\documentclass[twocolumn,pra,amssymb,amsmath,natbib,showpacs,floatfix,%
superscriptaddress]{revtex4}%

\usepackage[dvips]{graphicx}
\usepackage{color}

\newcommand{\rmi}{{\rm i}}
\newcommand{\rmd}{{\rm d}}

\begin{document}
\title{ Probing two-particle exchange processes in two-mode Bose-Einstein condensates}

\author{Luis Benet}%
\email{benet@fis.unam.mx}
\affiliation{Instituto de Ciencias F\'{\i}sicas, Universidad Nacional Aut\'onoma de M\'exico (UNAM)\\
Apdo. Postal 48--3, 62251 Cuernavaca, M\'exico}%
\author{Diego Espitia}
\affiliation{Instituto de Ciencias F\'{\i}sicas, Universidad Nacional Aut\'onoma de M\'exico (UNAM)\\
Apdo. Postal 48--3, 62251 Cuernavaca, M\'exico}%
\author{Daniel Sahag\'un}%
\affiliation{Instituto de F\'isica, Universidad Nacional Aut\'onoma de M\'exico (UNAM)\\
  Apartado Postal 20-364, 01000 Cd. Mx., M\'exico}%

\date{\today}

\begin{abstract}
We study the fidelity decay and its freeze for an initial coherent state
of two-mode Bose-Einstein condensates in the Fock regime considering a Bose-Hubbard model
that includes two-particle tunneling terms. By using linear-response theory we find scaling
properties of the fidelity as a function of the particle number that prove the existence of
two-particle mode-exchange when a non-degeneracy condition is fulfilled. Tuning the energy
difference of the two modes serves to distinguish the presence of two-particle mode-exchange
terms through the appearance of certain singularities.
We present numerical calculations that illustrate our findings, and
propose exploiting a Feshbach resonance to verify experimentally our predictions.
\end{abstract}


\maketitle


\section{Introduction}

The Bose-Hubbard model became a workhorse to describe interactions of ultracold bosonic gases
trapped by neighboring potentials after its striking success with the Mott insulator-superfluid
transition ~\cite{JakshEtAl1998,GreinerEtAl2002}. Its most simple physical realization, when only two
bosonic states can be occupied~\cite{MilburnEtAl1997}, is experimentally obtained by trapping the
condensate in a double-well potential~\cite{Gati2007}. This system is interesting because it is
the simplest scheme for atom interferometry. In addition to interference
phenomena~\cite{AndrewsEtAl97,GrossEtAl2010}, it also exhibits quantum tunneling and self trapping
effects~\cite{AndersonKasevich1997,AlbiezEtAl2005} such as Josephson
oscillations~\cite{CataliottiEtAl2001}. It has even been used to produce and study many-particle
entanglement~\cite{MicheliEtAl2003} and dynamically generate spin-squeezed
states~\cite{JuliaDiazEtAl2012}. Alternative methods to optical lattices have been demonstrated by
splitting a single-component Bose-Einstein condensate (BEC) on atom chips either with pure DC
magnetic fields~\cite{gunther2007} or by dressing static fields with RF
potentials~\cite{SchummEtAL2005, joEtAl2007, BauEtAl2010} as proposed in~\cite{ZobayEtAl2001}.
Understanding the effects originated on inter-atomic collisions has already been exploited to
overpass the classical limit in atom interferometers~\cite{GrossEtAl2010}, for example. Here we
employ an extended Bose-Hubbard Hamiltonian to increase the possibilities in this direction.

Several interchange terms arise in the Bose-Hubbard model in a double-well potential where only
the lowest level in each well is
populated and the corresponding wave functions have a small
overlap~\cite{Ananikian2006,Gati2007}. In particular, two terms accounting for two-particle
mode-exchange processes appear in the derivation of the Hamiltonian. These terms are often
neglected assuming that two-particle processes are rare for diluted ultracold gases.
Yet, Ref.~\cite{Ananikian2006} points out that there is a better agreement with the experimental
results when these processes are included. In this paper we probe the relevance of these terms
by studying dynamical properties linked with two-particle tunneling processes. We consider the
dynamical stability of the quantum evolution under small system perturbations for the two-mode
Bose-Hubbard model using the fidelity or Loschmidt echo~\cite{Peres1984,Jalabert2001, Gorin2006},
whose decay has been studied for different parameter ranges and types of perturbations in the
Bose-Hubbard model~\cite{Bodyfelt2007,Manfredi2008,Zheng2009}.

Prosen and \v{Z}nidari\v{c} noticed that fidelity stops decaying, staying essentially constant
(modulo some
oscillations) for relatively long times whenever the time-averaged expectation value of the
perturbation vanishes~\cite{ProsenZnid2003}. This phenomenon is called fidelity freeze. It was
later shown that symmetries can also induce this behavior if the diagonal matrix elements of
the perturbation vanish, e.g. when the perturbation is not invariant under the time-reversal
symmetry~\cite{GorinEtAl2006}. Note that the freeze of fidelity was actually observed in
simulations for bosonic and fermionic many-body systems~\cite{Manfredi2008,Manfredi2006}, but
it was attributed to the non-linearities introduced by the interactions between the particles.
Our purpose is to draw attention to this phenomenon and exploit it, within the context of the
Bose-Hubbard model.

The paper is organized as follows. In Sect.~\ref{sect:FidFreeze}, from a generalized
two-mode Bose-Hubbard Hamiltonian we derive a the fidelity freeze $F_{Fr}$
associated to an initial macroscopic trial state~\cite{Amico1998}. In
Sect.~\ref{sect:scaling}
we show analytically and numerically that the scaling properties of the
fidelity freeze display a transition
in terms of the number of particles if the interaction includes two-particle mode-exchange
terms. In addition, when certain degeneracy condition is fulfilled by tuning the energy
difference of the two levels, the fidelity freeze tends abruptly to zero. This yields insight
into many-body tunneling processes and provides a method to calibrate the system to enhance
the fidelity freeze. In Sect.~\ref{sect:concl} we summarize our results and address
the possibility to access the parameter range of interest in this paper
considering ${}^{87}$Rb and ${}^{85}$Rb.

\section{Fidelity freeze for the two-mode Bose-Hubbard model}
\label{sect:FidFreeze}

The fidelity amplitude is the overlap of the time-evolution of an initial state under a
reference interaction ${\hat H}_0$ with the evolution of the same initial state under a
slightly different Hamiltonian ${\hat H} = {\hat H}_0 + \lambda \hat{V}$~\cite{Gorin2006}:
\begin{equation}
\label{eq:fidAmp}
f(t) = \langle \Psi_0 | \hat{U}_0(-t) \hat{U}_\lambda(t) | \Psi_0 \rangle .
\end{equation}
Here, $| \Psi_0 \rangle$ is the initial state under consideration,
$\hat{U}_0(t) = \hat{T} \exp[- \rmi \, {\hat H_0} t / \hbar]$ is
the (time-ordered) unitary time-evolution associated to the reference Hamiltonian,
$\hat{U}_\lambda(t)$ is the corresponding time-evolution of the perturbed
Hamiltonian, and the perturbation strength is denoted formally by $\lambda$. The modulus
squared of the fidelity amplitude, $F(t) = | f(t)|^2$,
is the fidelity or Loschmidt echo~\cite{Peres1984,Jalabert2001}. Clearly, $F(t)$ is a
measure of the sensitivity of the time evolution of $|\Psi_0\rangle$ to system
perturbations. Another interpretation is that of an echo: $|\Psi_0\rangle$ evolves under
${\hat H_0}$ up to time $t$, when the system is suddenly reversed with respect
to time, and then evolves under the action of ${\hat H}$; the Loschmidt echo
compares the whole evolution with the initial state, thus quantifying the degree of
irreversibility of the system. The operator
${\hat M}_\lambda(t) = \hat{U}_\lambda(-t) \hat{U}_0(t)$ is referred as the echo
operator.

We consider the generalized Bose-Hubbard model $\hat{H}_{\rm BH} = \hat{H}_0+\hat{V}$
defined by
\begin{flalign}
\label{eq:H0}
\hat{H}_0 & = \epsilon_1 \hat{n}_1 + \epsilon_2 \hat{n}_2 +
      \frac{U}{2}\big[ \hat{n}_1(\hat{n}_1 - 1)+\hat{n}_2(\hat{n}_2 - 1) \big], \\
\label{eq:V}
\hat{V} & = -J_1 \big(\hat{b}_1^\dagger \hat{b}_2 + \hat{b}_2^\dagger \hat{b}_1\big)
      - \frac{J_2}{2} \big[ (\hat{b}_1^\dagger)^2 \,\hat{b}_2^2 +
      (\hat{b}_2^\dagger)^2 \,\hat{b}_1^2 \big].\quad
\end{flalign}
As usual, $\hat{n}_i = \hat{b}_i^\dagger \hat{b}_i$ ($i=1,2$) is the particle number
operator of the $i$th-mode, with $\hat{b}_i^\dagger$ and $\hat{b}_i$ the corresponding
bosonic creation and annihilation operators, respectively. The single-particle energies
of each mode are denoted by $\epsilon_i$, $U$ is the two-particle on-site interaction,
$J_1$ is the energy of the usual (one-particle) Josephson tunneling or mode-exchange term,
and $J_2$ is the energy associated to two-particle tunneling processes that we probe here. The
total number of particles $n=n_1+n_2$ is a conserved quantity; fixing $n$, the Hilbert-space
dimension is simply $n+1$. The Hamiltonian $\hat{H}_{\rm BH}$ defined through Eqs.~(\ref{eq:H0})
and (\ref{eq:V}) is a generalization of the usual two-mode approximation used to describe the
bosonic Josephson junction~\cite{MilburnEtAl1997,Gati2007}. Below we use $J_1$
and $J_2$ as perturbation parameters, replacing $\lambda$ in Eq.~(\ref{eq:fidAmp}).

We are interested in the so-called Fock regime ($U \gg J_1 > J_2$)
because the fidelity freeze can be observed there.
We choose the  Fock (occupation-number)  basis  defined by
$| \mu_1, \mu_2 \rangle = ({\mu_1!\,\mu_2!})^{-1/2} (\hat{b}_1^\dagger)^{\mu_1}
(\hat{b}_2^\dagger)^{\mu_2} |0\rangle$, where $|0\rangle$ is the vacuum state; since
$n=\mu_1+\mu_2$ is conserved, we use the short-hand notation
$| \mu_1\rangle \equiv |\mu_1,\mu_2\rangle$. By definition, $\hat{H}_0$ is diagonal in the Fock
basis and $\hat{V}$ has vanishing diagonal matrix elements. Then, considering $\hat{H}_0$ as the
reference interaction and $\hat{V}$ as the perturbation or residual interaction, the conditions
to observe the fidelity freeze are fulfilled~\cite{GorinEtAl2006}. The unperturbed
spectrum is given by $E_\mu = E_0 + \mu (\epsilon_1-\epsilon_2 -U n) + \mu^2 U$ with
$E_0 = \epsilon_2 n + U n(n-1)/2$, where $\mu=0,\dots n+1$ labels the Fock states
by mode occupation; notice the parabolic shape of $E_\mu$ in terms of $\mu$ for
non-vanishing $U$. As it is often done we use the Heisenberg time
$t_H = 2\pi\hbar/\overline{d}$ as the unit of time, where $\overline{d}$ is the average level
spacing of the unperturbed Hamiltonian ${\hat H_0}$.

We compute the fidelity decay by noting that $\hat{M}_\lambda(t)$ is the
time-evolution propagator associated with the time-dependent Hamiltonian
$\hat{V}_{\rm I}(t)=\hat{U}_0(-t) \hat{V} \hat{U}_0 (t)$ in the interaction
picture~\cite{Prosen2002}. We use Dyson's series on the perturbation parameters $J_r$
($r=1,2$) truncated to the second order~\cite{Prosen2002,BHS11}. This approach is called
linear response theory.

We write the fidelity amplitude as $f(t) = 1 + f_1 + f_2 + {\cal O}(J_i^3)$, where the first-
and second-order corrections (in both $J_1$ and $J_2$) read
\begin{flalign}
  \label{eq:f1}
  f_1 = & \sum_r \sum_{\mu,\nu} A_\mu^* A_\nu V^{(r)}_{\mu,\nu} \,
    {\mathcal I}_1[t;\Omega_{\mu,\nu}],  \\
  \label{eq:f2}
  f_2 = & \sum_{r,s} \sum_{\mu,\nu,\rho} A_\mu^* A_\nu V^{(r)}_{\mu,\rho} V^{(s)}_{\rho,\nu} \,
    {\mathcal I}_2[t;\Omega_{\mu,\rho},\Omega_{\rho,\nu}] .
\end{flalign}
Here, the matrix elements of the perturbation in the interaction picture are
$\langle \mu | \hat{V}_{\rm I}(t) | \nu \rangle = \sum_r V_{\mu,\nu}^{(r)}
\exp[\rmi \Omega_{\mu,\nu} t ]$ with $\hbar \Omega_{\mu,\nu} = E_\mu-E_\nu$,
and $A_\mu$ are the expansion coefficients of the initial state in the Fock basis. In
Eqs.~(\ref{eq:f1}) and~(\ref{eq:f2}), greek letters represent the basis states and $r,s=1,2$
stand for the one- or two-particle tunneling terms of $\hat{H}_{\rm BH}$. These
matrix elements read
\begin{flalign}
  \label{eq:MatElem}
  V_{\mu,\nu}^{(r)} = J_r\langle \mu | V^{(r)} | \nu \rangle = J_r (g_{\mu,n-\nu}^{(r)}\delta_{\mu-r,\nu} +
  g_{n-\mu,\nu}^{(r)} \delta_{\mu, \nu-r}),
\end{flalign}
where $g_{\mu,\nu}^{(r)} = [ \binom{\mu}{r}\binom{\nu}{r} ]^{1/2}$.
To second order in the perturbations, the fidelity is
\begin{flalign}
  \label{eq:Fid}
    F(t) = 1 + 2\, \Re(f_1) + 2\, \Re(f_2) + |f_1|^2.
\end{flalign}

The time dependence of Eqs.~(\ref{eq:f1}) and~(\ref{eq:f2}) appears in the
(time-ordered) integrals ${\mathcal I}_p [t; \Omega_1,\dots,\Omega_p]$, where $p$ stands for the
order in the Dyson's series. These integrals can be expressed recursively as
\begin{flalign}
  \label{eq:IntegralR}
  {\mathcal I}_{p+1}&[t; \Omega_1,\dots,\Omega_{p+1}] = \nonumber\\
     = & -\frac{\rmi}{\hbar }
      \int_0^t \rmd t_1 \exp[\rmi \Omega_1 t_1] \,
              {\mathcal I}_p[t_1; \Omega_2,\dots,\Omega_{p+1}] ,
\end{flalign}
where ${\mathcal I}_0[t] = 1$ defines the initial value of the recursion. These integrals
produce terms that oscillate in time as long as the frequencies $\Omega_{\mu,\nu}$ appearing
in the exponentials do not vanish, i.e. when the unperturbed spectrum is non-degenerate. Yet,
certain frequency combinations may vanish and yield secular terms which grow at least linearly
in time. We assume that the unperturbed spectrum is non-degenerate, which can be assured
by choosing properly the energy difference of the two modes
$\Delta \epsilon = \epsilon_2-\epsilon_1$. Then, without the secular contributions, to
second-order the fidelity displays quasi-periodic oscillations in time; this is the freeze
of the fidelity. The freeze of the fidelity lasts as long as the second-order approximation
is valid; eventually, higher-order contributions dominate the evolution and secular terms
appear that destroy the freeze of the fidelity.

Equation~(\ref{eq:Fid}) is valid for any initial state. We consider as the initial state
a normalized macroscopic trial state of the
form~\cite{Amico1998}
\begin{flalign}
  | \Psi_0 \rangle & = (\alpha\hat{b}_1^\dagger + \beta \mathrm{e}^{\rmi\phi} \hat{b}_2^\dagger)^n
    |0 \rangle \nonumber\\
  \label{eq:Psi0}
    & = \sum_\mu
    \binom{n}{\mu}^{1/2} \alpha^\mu \beta^{n-\mu} \mathrm{e}^{\rmi(n-\mu)\phi} | \mu \rangle .
\end{flalign}
This initial state is coherent~\cite{Amico1998}; with
$\alpha=(n_1/n)^{1/2}$ and $\beta=(n_2/n)^{1/2}$, it
corresponds to the mean-field state having $n_1$ particles in the first mode and $n_2=n-n_1$
in the second one.

Inserting Eqs.~(\ref{eq:MatElem}) and~(\ref{eq:Psi0}) in~(\ref{eq:f1}) and~(\ref{eq:f2}),
we obtain
\begin{flalign}
\label{eq:f1td}
f_1 = &\sum_{r} J_r \sum_{\mu,\nu} A_\mu^* A_\nu
    \, {\cal I}_1[t;\Omega_{\mu,\nu}] \qquad \nonumber\\
\times & \Big[ g^{(r)}_{\mu,n-\nu} \delta_{\nu,\mu-r} +
    g^{(r)}_{n-\mu,\nu} \delta_{\nu,\mu+r} \Big], \\
\label{eq:f2td}
f_2 = & \sum_{r,s} J_r J_s \sum_{\mu,\nu,\rho} A_\mu^* A_\nu
    \, {\cal I}_2[t;\Omega_{\mu,\rho},\Omega_{\rho,\nu}] \nonumber\\
\times & \Big[ g^{(r)}_{\mu,n-\rho} \delta_{\rho,\mu-r}
    \big( g^{(s)}_{\rho,n-\nu} \delta_{\nu,\rho-s}
    + g^{(s)}_{n-\rho,\nu} \delta_{\nu,\rho-s} \big)\qquad \nonumber\\
+ & g^{(r)}_{n-\mu,\rho} \delta_{\rho,\mu+r}
    \big( g^{(s)}_{\rho,n-\nu} \delta_{\nu,\rho-s}
    + g^{(s)}_{n-\rho,\nu} \delta_{\nu,\rho+s}\big) \Big].
\end{flalign}
The time dependence can be further described by noting that ${\cal I}_1[t;\Omega] =
(1-\exp(\rmi \Omega t))/\hbar\Omega$, and ${\cal I}_2[t;\Omega_1, \Omega_2] =
({\cal I}_1[t;\Omega_1]-{\cal I}_1[t;\Omega_1+\Omega_2])/\hbar\Omega_2$, for
$\Omega_1$ and $\Omega_2$ non-zero. These conditions are fulfilled by the assumption
of a non-degenerate spectrum. Yet, for $\Omega_1+\Omega_2 = 0$ a secular term is
obtained for $f_2$ which has the form $-\rmi t/\hbar$. This
term does not affect the fidelity according to Eq.~(\ref{eq:Fid}), since it is
purely imaginary. Then, to second-order in the tunneling rates, the time-dependence of
the fidelity is at most quasi-periodic, hence fidelity exhibits a freeze.
The time during which the fidelity freeze lasts scales as the inverse of the
perturbation and the inverse of $n$; see~\cite{BHS11}.

In order to obtain the fidelity freeze we extract the time-independent
contributions of the integrals~(\ref{eq:IntegralR}) in the expression for
$\Re(f_1)$, $\Re(f_2)$ and $|f_1|^2$. Including the dependency of $A_\mu$
on the phase $\phi$, cf. Eq.~(\ref{eq:Psi0}), we obtain
\begin{flalign}
\label{eq:f1ti}
& \Re\Big[\mathrm{e}^{\rmi p \phi} \, {\cal I}_1[t;\Omega]\Big]
  \leadsto \frac{\cos(p\phi)}{\hbar \Omega}, \\
\label{eq:f2ti}
& \Re\Big[\mathrm{e}^{\rmi p \phi} \, {\cal I}_2[t;\Omega_1,\Omega_2] \Big]
  \leadsto \frac{\cos(p\phi)}{\hbar^2 \Omega_2}
      \Big(\frac{1}{\Omega_1}-\frac{1-\delta_{1,-2}}{\Omega_1+\Omega_2}\Big), \\
\label{eq:f1tisquare}
& \Re\Big[\mathrm{e}^{\rmi p \phi} {\cal I}_1^*[t;\Omega_1] \, {\cal I}_1[t;\Omega_2] \Big]
  \leadsto \frac{\cos(p\phi)}{\hbar^2\Omega_1 \Omega_2}(1+\delta_{1,2}).
\end{flalign}
Here, the right-hand side of these expressions are the time-independent
contributions, where $p$ is an integer related to the indexes of the Fock states involved,
and we have used the Kronecker-delta $\delta_{1,-2}$ to indicate that the frequencies
satisfy $\Omega_1= -\Omega_2$ (indexes are reversed), and $\delta_{1,2}$ to denote
that $\Omega_1=\Omega_2$ (indexes are the identical). Note that in Eq.~(\ref{eq:f2ti}) the
secular term related to $\Omega_1+\Omega_2 = 0$ is not included due to the $\delta_{1,-2}$.

Inserting Eqs.~(\ref{eq:f1td}) and~(\ref{eq:f2td}) into (\ref{eq:Fid}), and using the
time-independent contributions, Eqs.~(\ref{eq:f1ti}) to~(\ref{eq:f1tisquare}),
we obtain
\begin{widetext}
\begin{flalign}
\label{eq:FidFreeze}
F_{\rm Fr} = 1 +
  2 & \sum_{r,\mu} J_r \frac{ |A^*_\mu A_{\mu\pm r}| {\cal G}^{(\pm r)}_{\mu}
    \cos(r\phi)}{\hbar\Omega_{\mu,\mu\pm r}}
  + 2 \sum_{r,s,\mu} J_r J_s
    \frac{ |A^*_\mu A_{\mu\pm r}| {\cal G}^{(\pm r)}_{\mu}  {\cal G}^{(\pm s)}_{\mu\pm r}
      \cos((\mp r \mp s)\phi)}{\hbar^2 \Omega_{\mu\pm r, \mu\pm r \pm s}}
      \Big(\frac{1}{\Omega_{\mu,\mu\pm r}}
          -\frac{1-\delta_{\pm r, \mp s}}{\Omega_{\mu,\mu\pm r\pm s}}\Big) \nonumber\\
  + & \sum_{r,s,\mu,\nu} J_r J_s
    \frac{ |A^*_\mu A_{\mu\pm r}| \, |A^*_\nu A_{\nu\pm s}|
      {\cal G}^{(\pm r)}_{\mu} {\cal G}^{(\pm s)}_{\nu\pm r}
      \cos((\mp r \pm s)\phi)}{\hbar^2\Omega_{\mu,\mu\pm r}\Omega_{\nu, \nu\pm s}}
      (1+\delta_{\pm r, \pm s}).
\end{flalign}
\end{widetext}
The coefficients ${\cal G}^{(-r)}_{\mu} = g^{(r)}_{\mu,n-\mu+r}$ and
${\cal G}^{(+r)}_{\mu} = g^{(r)}_{n-\mu,\mu+r}$ are introduced to have a more compact expression.
The signs
of $r$ and $s$ are independent and correspond to the distinct possibilities imposed
by the Kronecker deltas that appear in Eqs.~(\ref{eq:f1td}) and~(\ref{eq:f2td}).
Equation~(\ref{eq:FidFreeze}) is a central result of this paper.

\begin{figure}
  \includegraphics[angle=0,width=8.8cm]{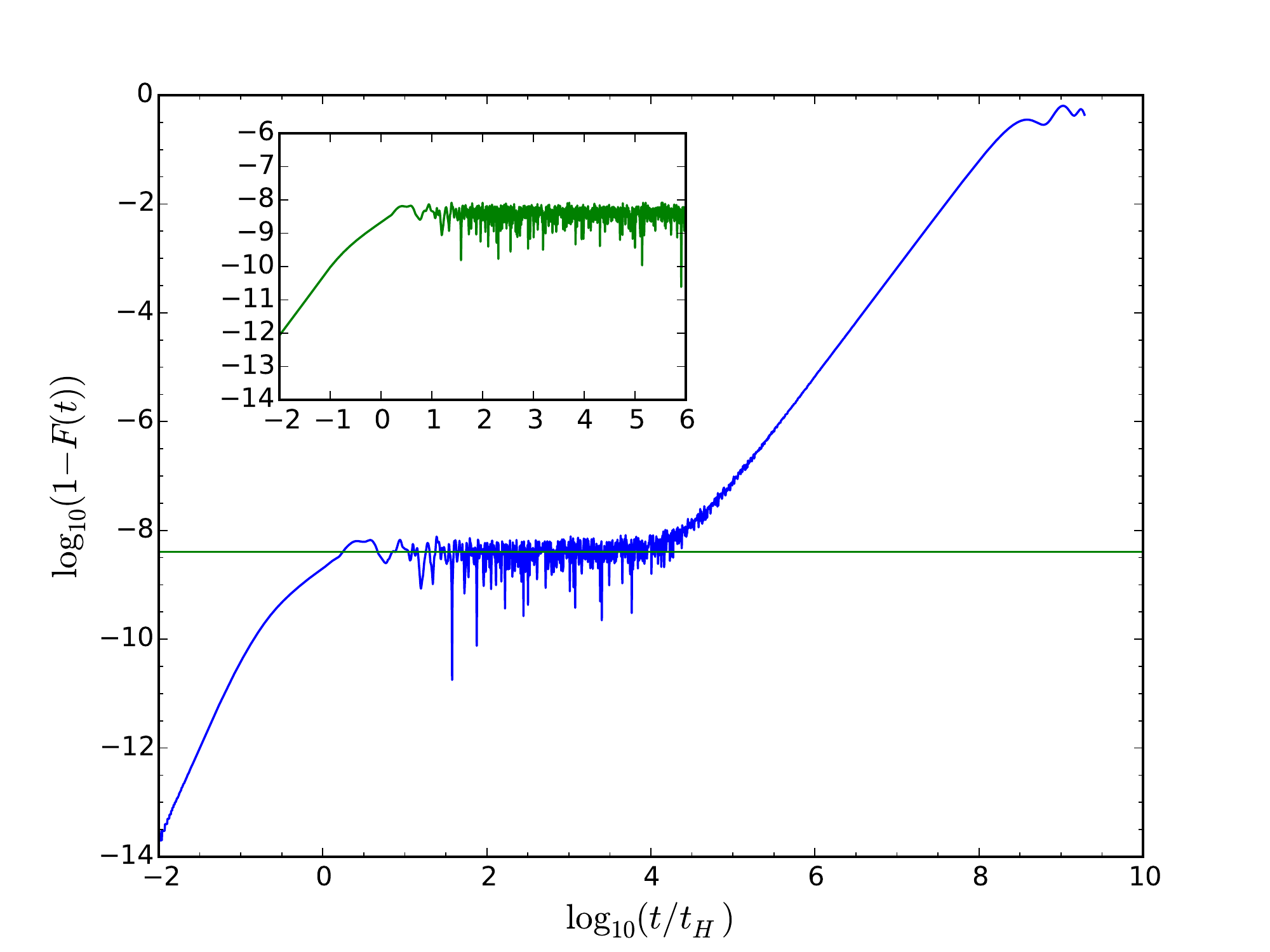}
  \caption{
  (Color online) Time-dependence (in Heisenberg time units) of the fidelity obtained
  numerically for the extended Bose-Hubbard model in log-log scale, using an
  initial coherent
  state Eq.~(\ref{eq:Psi0}) with $n_1=n_2=n/2=64$ and $\phi=\pi/4$.
  The parameters for the Hamiltonian are $U=1$, $J_1=10^{-6}$, $J_2=10^{-8}$,
  $\epsilon_1=0.76$ and $\epsilon_2=0.93$.
  The horizontal line corresponds to the value of the fidelity freeze $F_{\rm Fr}$
  obtained from Eq.~(\ref{eq:FidFreeze}). In the inset we present the result of the
  second-order linear-response theory.
  }
  \label{fig:Ffreeze}
\end{figure}

In Fig.~\ref{fig:Ffreeze} we show an example of the decay of fidelity for a coherent state
with $n_1=n_2=n/2$ and $\phi=\pi/4$ obtained numerically. The figure illustrates the oscillations
during the freeze of the fidelity, the eventual decay,
and the value obtained from
Eq.~(\ref{eq:FidFreeze}) for the freeze of the fidelity (horizontal green line). Time is
measured in Heisenberg-time units $t_H$. The parameters of the model are $U=1$, $J_1=10^{-6}$,
$J_2=10^{-8}$, $\epsilon_1=0.76$, $\epsilon_2=0.93$ and $n=128$; the values of $\epsilon_i$
assure the non-degeneracy of the spectrum of
$\hat{H}_0$ (see below). These parameters have been chosen to simplify the numerics;
other values display qualitative similar behavior as long as we are in the Fock
regime. In the inset we display the result considering the second-order
expansion~(\ref{eq:Fid}); the value of $F_{\rm Fr}$ is an average
over the quasi-periodic oscillations that take place during the freeze.

\begin{figure}
  \includegraphics[angle=0,width=8.8cm]{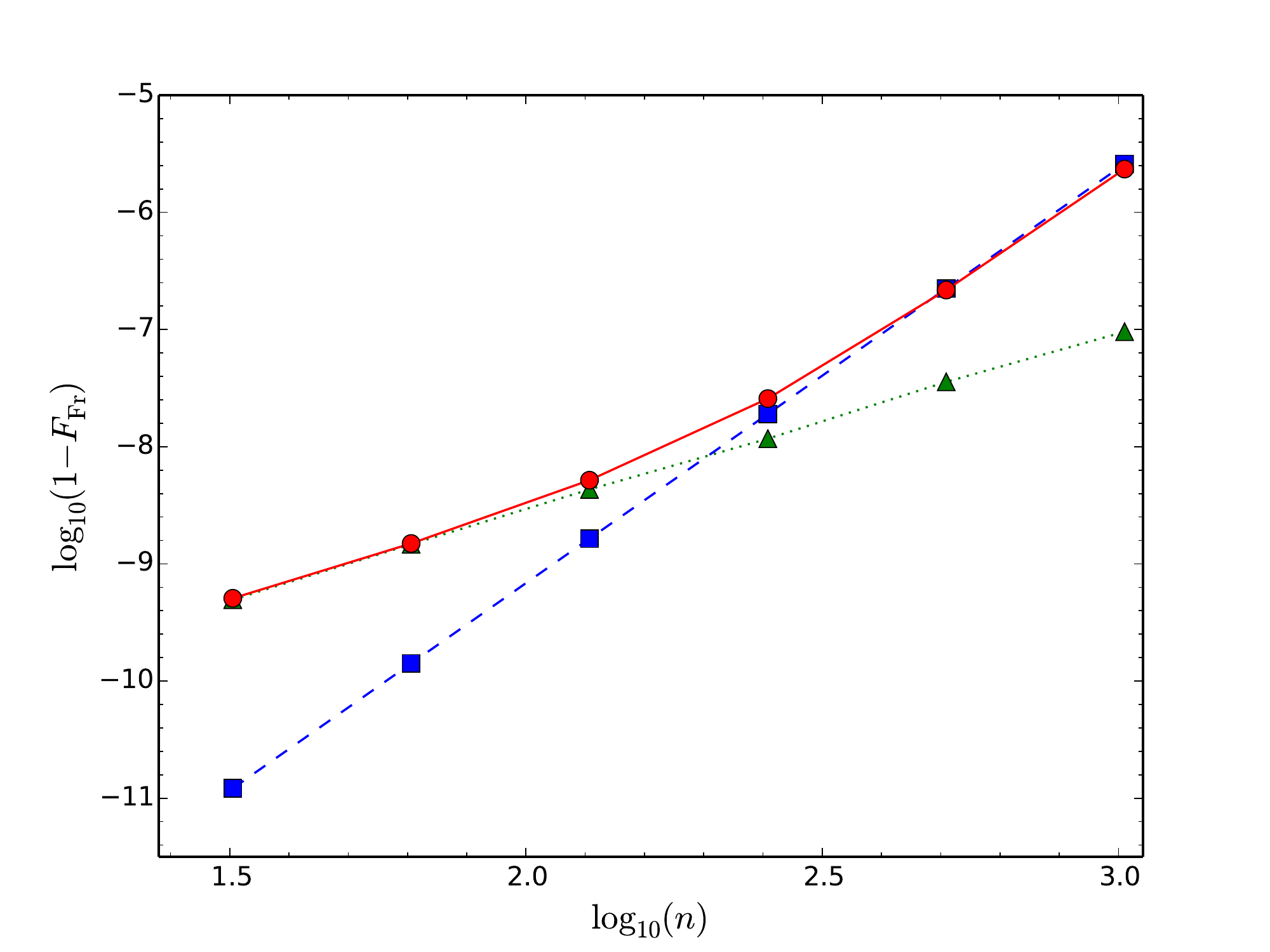}
  \caption{ (Color online) Log-log plot of $1-F_{\rm Fr}$ as a function of the number of
  particles, for an initial symmetric coherent state $n_1=n_2=n/2$.
  Triangles (green, dotted line) correspond
  to the parameters $J_1=10^{-6}$ and $J_2=0$, squares (blue, dashed line) to $J_1=0$ and
  $J_2=10^{-8}$, and circles (red, continuous line) to $J_1=10^{-6}$ and $J_2=10^{-8}$; the
  remaining parameters are those of Fig.~\ref{fig:Ffreeze}.
  }
  \label{fig2}
\end{figure}

\section{Scaling properties of the fidelity freeze}
\label{sect:scaling}

We address now the scaling of $F_{\rm Fr}$ in terms of the number of particles. An estimate
of the scaling properties is obtained considering the maximum contribution of the $n$-dependent
terms in Eq.~(\ref{eq:FidFreeze}). This follows from a Fock state that we write as
$\mu = \lambda n$, and then use Stirling's formula for large $n$. It can be shown that
$|A_\mu^* A_{\mu\pm r}|\sim n^{1/2}$ and ${\cal G}^{(\pm s)}_{\mu\pm r}\sim n^s$ for
$\lambda = \alpha^2$. The scaling laws of the time-independent contributions thus read
$\Re(f_1)\sim J_r n^{r-1/2}$, $\Re(f_2)\sim J_r^2 n^{2r-1/2}$ and
$\Re(|f_1|^2)\sim J_r^2 n^{2r-1}$.
Hence, the asymptotic dominating contribution for the fidelity freeze scales as
\begin{flalign}
\label{eq:scale}
1-F_{\rm Fr} \sim J_r^2 n^{2r-1/2} .
\end{flalign}
This result predicts a different scaling for each of the tunneling terms $J_1$ and $J_2$. Thus,
$F_{\rm Fr}$ exhibits a transition from a behavior dominated by $J_1$ to a regime where
$J_2$ dominates, around $n\sim J_1/J_2$. Figure~\ref{fig2} is the numerical confirmation
of this statement. The data points were obtained numerically from the time series (cf.
Fig.~\ref{fig:Ffreeze}), using the maxima of the quasi-periodic oscillations of $1-F(t)$
during the freeze; these values underestimate the theoretical expectation for $F_{\rm Fr}$.
Fitting the data to straight-lines when either $J_2$ or $J_1$ are absent yields the
slopes $1.52$ and $3.54$, respectively. These values are in excellent agreement with the
$3/2$ and $7/2$ predicted by Eq.~(\ref{eq:scale}), thus showing that the scaling properties
of the fidelity freeze in terms of $n$ probe the presence of two-particle tunneling
processes. Equation~(\ref{eq:scale}) remains valid for small departures from the
symmetric initial coherent states, i.e., $\lambda\sim \alpha^2$.

An important assumption that we made in the derivation of Eqs.~(\ref{eq:FidFreeze})
and~(\ref{eq:scale}) is that the spectrum of $\hat{H}_0$ is non-degenerate, which can
be fulfilled by tuning $\Delta \epsilon$, the energy difference of the two modes. As we approach
a degeneracy, the appearance of secular terms makes Eq.~(\ref{eq:FidFreeze}) no longer valid.
This can be exploited to probe the relevance of two-mode exchange processes.

\begin{figure}
  \includegraphics[angle=0,width=9.5cm,trim=3cm 1cm 0 2cm, clip=true]{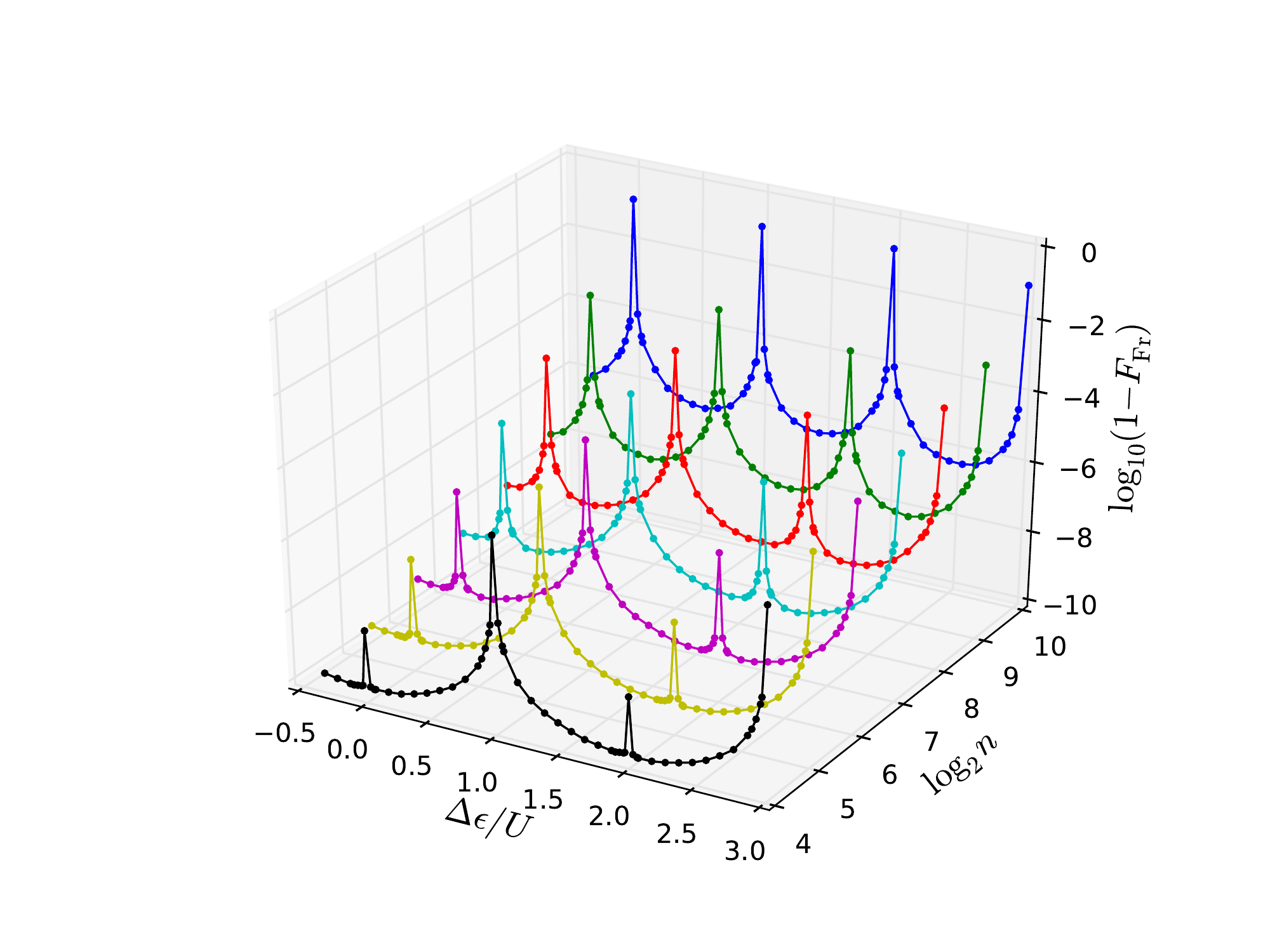}
  \caption{ (Color online) Behavior of $\log_{10}(1-F_{Fr})$ as a function of the energy
  difference between modes $\Delta\epsilon$ scaled by the two-particle interaction coefficient
  $U$ of the Bose-Hubbard model, Eq. \ref{eq:H0}. The 3D plot depict the appearance of a
  peak around $\Delta\epsilon/U=0,2$ which becomes noticeable as the particle number
  $n$ increases.
  }
  \label{fig:scaling}
\end{figure}

To clarify this idea we consider the Fock state $\mu_0 = \lfloor (n+\Delta\epsilon/U)/2\rceil$
whose energy is the minimum of the spectrum of ${\hat H_0}$, where $\lfloor x \rceil$ is the
round-to-nearest integer function; note that this minimum corresponds to the parabolic shape
of $E_\mu$ induced by a non-vanishing $U$. Assuming that $n$ is even for concreteness, it can be shown that
$\Delta\epsilon/U=0$ implies that $E_{\mu_0-1} = E_{\mu_0+1}$, meaning that the Fock states
$\mu_0-1$ and $\mu_0+1$ are degenerate; these states are coupled by a two-mode tunneling term.
The same holds for $\Delta\epsilon/U=2$, though the actual value of $\mu_0$ has
changed. For $\Delta\epsilon/U=1$ we have $E_{\mu_0} = E_{\mu_0+1}$, i.e. the ground state is
degenerate, which also holds for $\Delta\epsilon/U=3$; in this case, the states are coupled by
a one-particle tunneling term. Then, by tuning the single-particle energies, as we approach
$\Delta\epsilon/U=\pm 1$ or $\pm 3$, a peak in $\log_{10}(1-F_{\rm Fr})$ develops indicating
that the perturbation does
contain a one-particle tunneling term; likewise, a peak at $\Delta\epsilon/U=0$ and
$\pm 2$ appears
if there are two-particle tunneling processes. This is illustrated in
Fig.~\ref{fig:scaling}, which depicts $\log_{10}(1-F_{\rm Fr})$ in terms of $\Delta\epsilon/U$
for various even values of $n$. Note that the narrow peaks at $\Delta\epsilon/U=0,\pm 2$,
the signature of the two-particle tunneling,
grow for increasing values of $n$. For odd values of $n$ the same argument applies,
exchanging only the location of the peaks. Thus, by increasing $n$, the peaks associated
with the two-particle tunneling processes become comparable to those
associated to the one-particle tunneling processes; for big enough $n$ the distance
between prominent neighboring peaks is halved. This result means that the fidelity
freeze $F_{\rm Fr}$ can also be maximized by tuning $\Delta\epsilon/U$.

\section{Summary and outlook}
\label{sect:concl}

Summarizing, we have found that the fidelity freeze from an initial symmetric
coherent state is a sensitive quantity to two particle mode-exchange
processes in the Bose-Hubbard model. This sensitivity can be controlled with two
experimental parameters: the total atom number $n$ and the energy difference
between modes $\Delta\epsilon/U$. In terms of $n$, the fidelity freeze 
displays a
transition from a regime dominated by the one-particle exchange term, for small
particle numbers, to the dominance of two-particle tunneling processes when $n$
is large enough ($n\sim J_1/J_2$). There, the fidelity freeze can also be
maximized by tuning $\Delta\epsilon/U$.

Our findings hold in the Fock regime of a double well potential, i.e.
for $J_r/U \ll 1$.
Their test would face two technological challenges: measuring fidelity
and producing a BEC with an adequate atom number and confining geometry.
Measuring fidelity is not a simple task though it has been achieved in NMR
polarization echo-spin experiments~\cite{Levstein1998,Zangara2016} and in
periodically-kicked cold atoms~\cite{Wu2009}. Echo spectroscopy experiments
in cold atoms by Andersen et al.~\cite{Andersen2006} demonstrated measurements
of a quantum fidelity defined differently; theoretical aspects of that
definition are discussed in Ref.~\cite{Goussev2010}. Fidelity has
not yet been measured in two-mode Bose-Einstein condensates, though it has
been proposed in the context of cold optical lattices
Ref.~\cite{cucchietti2010}.

\begin{figure}
  \includegraphics[angle=0,width=8cm]{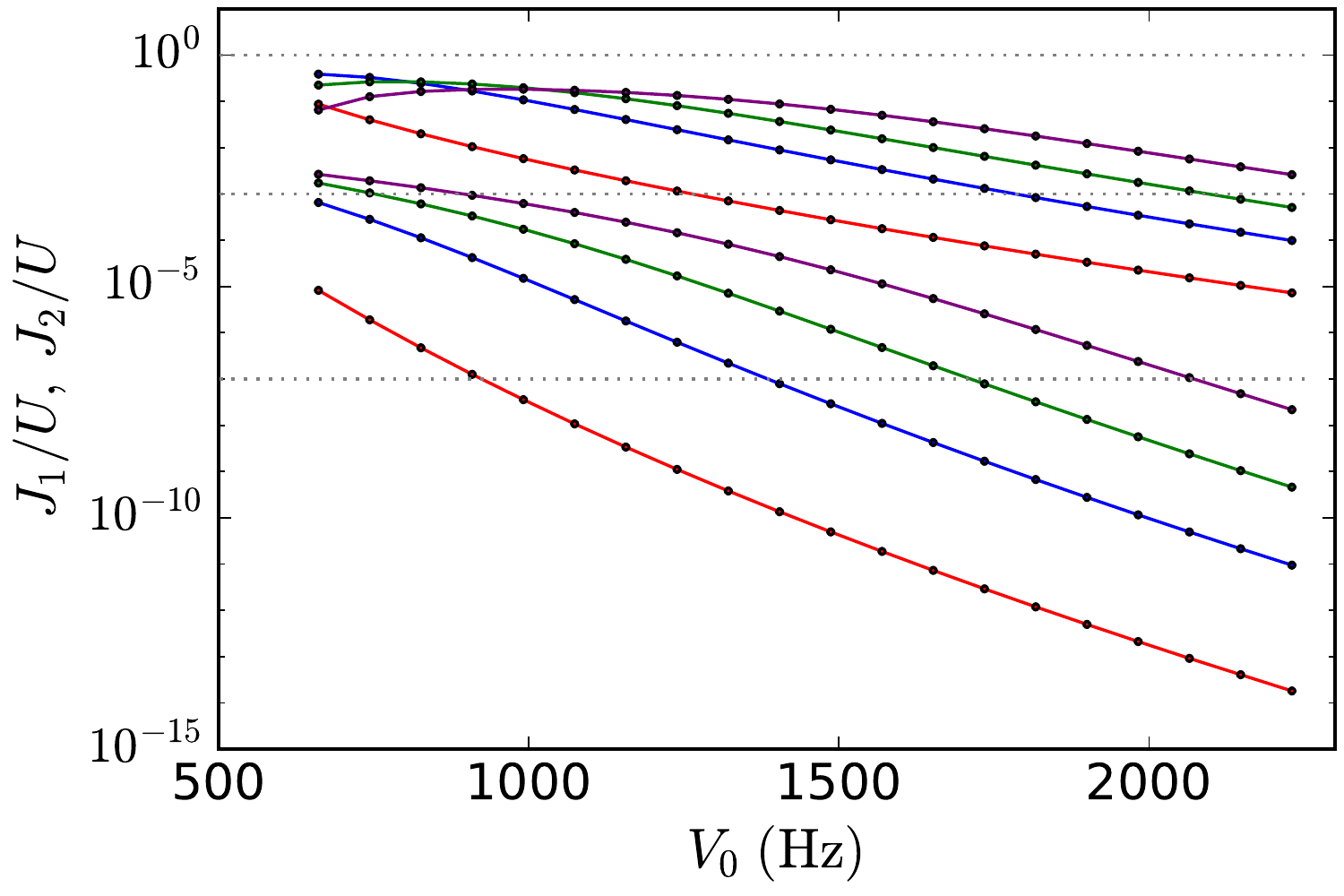}
  \caption{ Computed tunneling parameters of the two-mode BEC model 
  as a function of the potential-barrier height,
  where the trap considered corresponds to that of Ref.~\cite{AlbiezEtAl2005}.
  The red curves correspond to ${}^{87}$Rb, and the blue, green and
  purple to ${}^{85}$Rb with s-wave scattering lengths of $20a$, $50a$
  and $100a$, respectively, where $a=5.32$nm is the s-wave scattering
  length of ${}^{87}$Rb.
  The upper curves correspond to $J_1/U$ and the lower to $J_2/U$.
  }
  \label{fig4}
\end{figure}

Regarding what is an adequate BEC species to probe our results,
special care is required for having a long enough confinement
with $J_2$ big enough, so the effects addressed here can be observed
experimentally.
As an example, consider ${}^{87}$Rb using the same trap frequencies and
separation of the wells as in Ref.~\cite{AlbiezEtAl2005},
and vary the
potential height in order to reach the Fock regime. In this case, the
Heisenberg time is $\sim 5-7$~ms for 1500 bosons. The duration of the fidelity
freeze is $t_H \times 10^4$ (Fig.~\ref{fig:Ffreeze}), which is well inside normal
experimental times for Bose-Einstein condensates under ultra-high vacuum.

In Fig.~\ref{fig4}, we present the parameters
$J_1/U$ (upper curves) and $J_2/U$ (lower curve) using the formulae
of Ref.~\cite{Ananikian2006}, as a function of the barrier height $V_0$ for
$n=1500$ bosons. These results were obtained by integrating
the non-polynomial non-linear Schr\"odinger equation~\cite{Salasnich2002}
using the standard split-slit Fourier method~\cite{deVries1986,GatiPhD2007}.
The results show that for ${}^{87}$Rb (red curves), the two-particle
tunneling parameter $J_2/U$ is perhaps too small to yield any measurable
signal. For
instance, for $V_0\simeq 1500$Hz, where $J_1/U\sim 10^{-4}-10^{-3}$, we
obtain $J_2/U\sim 10^{-11}$. Increasing further the potential height
makes $J_2/U$ to decrease even further.

A more promising possibility is to consider other mechanisms that
increase $J_2/U$, e.g. approaching a Feshbach resonance; an obvious
candidate is ${}^{85}$Rb~\cite{cornish200}. Considering
the same parameters for the trap used above, the blue, green and purple
curves in Fig.~\ref{fig4} correspond to s-wave scattering length values of
$20a$, $50a$ and $100a$, respectively. Here we use $a=5.32$~nm, the s-save
scattering length of ${}^{87}$Rb, as unit to ease the comparisons. The results
show that larger values of $J_2/U$ are obtained and, in that sense, may
be accessible to experimental observation. Yet, we note that increasing the
s-wave scattering length by such amounts leads to three-body collisions which
have not been taken into account in our calculations.

Double-well potentials are now a common scenario for atom interferometry with
matter waves. Our findings could be useful to study coherence and decoherence
effects in this context. For instance, they could help to minimize decoherence
on interferometers using dense atomic clouds, where non-linearities due to
collisions can be exploited to improve their accuracy~\cite{GrossEtAl2010}, or
give rise to optimal methods for analyzing the interference fringes imprinted
by small energy differences between matter waves~\cite{BauEtAl2010}.

We are thankful to Wolf von Klitzing for discussions and encouragement. We acknowledge
financial support from DGAPA-PAPIIT (UNAM) projects IG-100616 and IA-103216,
and from CONACyT via the National Laboratory project 232652. LB acknowledges 
support from a Moshinsky fellowship (2012).


\end{document}